\documentclass{article}
\usepackage{multirow}
\usepackage{amsmath}
\usepackage{amssymb}
\usepackage{graphicx}
\usepackage{hyperref}
\usepackage{enumitem}

\title{QuantU-Net: Efficient Wearable Medical Imaging Using Bitwidth as a Trainable Parameter}
\author{Christiaan Boerkamp \and Akhil John Thomas}

\date{ VLV Medical}

\begin{document}

\maketitle

\begin{abstract}
Medical image segmentation, particularly tumor segmentation, is a critical task in medical imaging, with U-Net being a widely adopted convolutional neural network (CNN) architecture for this purpose. However, U-Net's high computational and memory requirements pose challenges for deployment on resource-constrained devices such as wearable medical systems. This paper addresses these challenges by introducing QuantU-Net, a quantized version of U-Net optimized for efficient deployment on low-power devices like Field-Programmable Gate Arrays (FPGAs). Using Brevitas \cite{pappalardo2023brevitas}, a PyTorch library for quantization-aware training, we quantize the U-Net model, reducing its precision to an average of 4.24 bits while maintaining a validation accuracy of 94.25\%, only 1.89\% lower than the floating-point baseline. The quantized model achieves an ~8x reduction in size, making it suitable for real-time applications in wearable medical devices. We employ a custom loss function that combines Binary Cross-Entropy (BCE) Loss, Dice Loss, and a bitwidth Loss function to optimize both segmentation accuracy and the size of the model. Using this custom loss function we have significantly cut down on the training time to find an optimal combination of bitwidth of the model and accuracy from a hypothetical \(6^{23}\) number of training sessions to a single training session by the aforementioned combination of loss functions.  The model's usage of integer arithmetic highlights its potential for deployment on FPGAs and other designated AI accelerator hardware. This work advances the field of medical image segmentation by enabling the deployment of deep learning models on resource-constrained devices, paving the way for real-time, low-power diagnostic solutions in wearable healthcare applications.
\end{abstract}

\section{Introduction}

Medical image segmentation, particularly tumor segmentation, is a crucial task in the field of medical imaging. U-Net, a convolutional neural network (CNN) architecture, has been widely adopted for this purpose due to its effectiveness in handling biomedical images. However, the high computational and memory requirements of U-Net pose challenges for deployment on resource-constrained devices such as mobile phones or embedded systems.

Quantization, the process of reducing the precision of the weights and activations of a neural network, offers a promising solution to this problem. By quantizing the model, we can significantly reduce its size and computational requirements, making it more suitable for deployment in real-world scenarios, particularly in wearable medical devices.

Wearable medical devices are increasingly being used for continuous health monitoring, early diagnosis and therapeutics.\cite{Guk2019Evolution, Babu2023Wearable, Yetisen2018Wearables} These devices often operate in environments where power usage is paramount, as a user's life can depend on the device functioning reliably over extended periods. For instance, A wearable device capable of real-time cancer diagnosis, such as an AI-assisted ultrasound scanner for early tumor detection, could provide users with immediate insights that might otherwise require clinical imaging appointments. If such a device fails, it could delay critical diagnoses, potentially leading to worsened prognoses or missed treatment opportunities. However, the computational demands of deep learning models like U-Net can be a significant barrier to their effective deployment in such devices.

The primary motivation for quantization in this study is to enable the deployment of the U-Net model on embedded devices, specifically Field-Programmable Gate Arrays (FPGAs) or other types of neural network accelerators. FPGAs are highly efficient for running quantized models due to their ability to perform integer arithmetic operations more efficiently than floating-point operations. According to studies, integer operations on FPGAs can be improved by an order of magnitude and require significantly less memory compared to floating-point operations \cite{giefers2016measuring, umuroglu2017finn, Yetisen2018Wearables}. This efficiency is crucial for wearable medical devices, where power consumption and battery life are critical factors.

By quantizing the U-Net model, we aim to reduce its computational and memory footprint, making it feasible to run on low-power embedded systems without compromising the accuracy of tumor segmentation. This approach not only enhances the efficiency of the device but also extends its battery life, ensuring reliable operation over longer durations.

The main contributions of this paper are as follows:

\begin{itemize}
    \item \textbf{Baseline Floating-Point U-Net}: We establish a baseline using a floating-point U-Net model for tumor segmentation.
    
    \item \textbf{Quantized U-Net with Brevitas}: We quantize the entire U-Net model using Brevitas\cite{pappalardo2023brevitas}, a PyTorch library for quantization-aware training.
    
    \item \textbf{Bitwidth Optimization}: We make the bitwidth a trainable parameter and use gradient descent combined with a custom loss function to find the smallest possible bitwidth that retains the accuracy of the floating-point model.

\end{itemize}

\section{Related Works}

Quantization-aware training (QAT) has been extensively studied as a method to mitigate the accuracy loss typically associated with quantization. QAT involves simulating the effects of quantization during the training process, allowing the model to adapt to the lower precision. Several studies have demonstrated the effectiveness of QAT in maintaining model accuracy while significantly reducing the model size and computational requirements. \cite{Fan2020Training}

Recent advances in QAT have explored the idea of treating bitwidth \cite{Chen2020A, Peters2023QBitOpt:} as a trainable parameter, allowing the model to dynamically adjust the precision of its weights and activations during training. This approach can lead to more efficient models by optimizing the bitwidth for different layers or even individual neurons. Quantization has also been explored in the context of medical imaging, where the computational demands of deep learning models can be particularly challenging.

\section{Dataset}

The dataset used in this study is the Breast Ultrasound Images Dataset \cite{aryashah2k_breast_ultrasound_images_dataset, al-dhabyani2020dataset}, a publicly available dataset. This dataset consists of breast ultrasound images categorized into three classes: normal, benign, and malignant. It is specifically designed for tasks such as tumor segmentation and classification, making it highly suitable for evaluating the performance of the U-Net architecture in medical image analysis.

\subsection{Dataset Description}

The dataset contains a total of 780 images, distributed across the three classes as follows:

\begin{itemize}
    \item \textbf{Normal}: Images showing no signs of tumors or abnormalities.
    \item \textbf{Benign}: Images containing benign tumors, which are non-cancerous and typically less aggressive.
    \item \textbf{Malignant}: Images containing malignant tumors, which are cancerous and require immediate medical attention.
\end{itemize}

Each image is provided in PNG format and is accompanied by a corresponding mask that outlines the region of interest (ROI), such as the tumor area. These masks are binary images where the tumor region is marked in white, and the background is black. The dataset also includes metadata, such as patient information and tumor characteristics, which can be used for further analysis.

\subsection{Dataset Preprocessing}

The dataset is loaded from a structured directory, with images and their corresponding masks organized into subdirectories for each category: benign, malignant, and normal. The grayscale images are resized to a fixed resolution of \(128 \times 128\) pixels, and their pixel values, along with those of the masks, are normalized to the range \([0, 1]\) by dividing by 255 to make them suitable for neural network input. For images without corresponding masks (e.g., normal cases), blank masks are generated to maintain consistency, and each image-mask pair is assigned a label (0 for benign, 1 for malignant, and 2 for normal). The dataset is then split into training (70\%), validation (15\%), and test (15\%) subsets using a stratified approach to preserve the class distribution across all subsets, ensuring a balanced representation of benign, malignant, and normal cases. Finally, the images and masks are transformed into PyTorch tensors with a data type of float32, and a channel dimension is added, resulting in a shape of \([1, 128, 128]\) for grayscale images, making them ready for deep learning model training.

\subsection{Dataset Challenges}

The dataset presents several challenges that are typical of medical imaging tasks:

\begin{itemize}
    \item \textbf{Class Imbalance}: The distribution of images across the three classes (normal, benign, malignant) is not uniform. This imbalance is addressed using techniques such as oversampling the minority classes or applying class-weighted loss functions during training.
    \item \textbf{Noise and Artifacts}: Ultrasound images often contain noise and artifacts due to the nature of the imaging modality. Preprocessing steps, such as normalization and augmentation, help mitigate these issues.
    \item \textbf{Small Dataset Size}: With only 780 images, the dataset is relatively small for deep learning tasks. Data augmentation and transfer learning are employed to overcome this limitation.
\end{itemize}

\section{Baseline U-Net Model}

The baseline model used in this study is a U-Net architecture \cite{ronneberger2015u}, a convolutional neural network (CNN) specifically designed for biomedical image segmentation. The U-Net architecture is characterized by its encoder-decoder structure with skip connections, which enables it to effectively capture both local and global features in medical images. The architecture consists of an encoder, a bottleneck, and a decoder, followed by a single output head for segmentation. Table~\ref{tab:unet_architecture} provides an overview of the U-Net architecture used in this study.

\begin{table}[ht]
\centering
\caption{Overview of the U-Net Architecture}
\label{tab:unet_architecture}
\resizebox{\textwidth}{!}{%
\begin{tabular}{|l|l|l|l|}
\hline
\textbf{Component} & \textbf{Layer} & \textbf{Feature Maps} & \textbf{Operation} \\ \hline
\multirow{4}{*}{Encoder} & ConvBlock 1 & 64 & Conv2d (3x3), ReLU, Conv2d (3x3), ReLU \\ \cline{2-4}
                          & MaxPool 1   & 64 & MaxPool2d (2x2, stride=2) \\ \cline{2-4}
                          & ConvBlock 2 & 128 & Conv2d (3x3), ReLU, Conv2d (3x3), ReLU \\ \cline{2-4}
                          & MaxPool 2   & 128 & MaxPool2d (2x2, stride=2) \\ \cline{2-4}
                          & ConvBlock 3 & 256 & Conv2d (3x3), ReLU, Conv2d (3x3), ReLU \\ \cline{2-4}
                          & MaxPool 3   & 256 & MaxPool2d (2x2, stride=2) \\ \cline{2-4}
                          & ConvBlock 4 & 512 & Conv2d (3x3), ReLU, Conv2d (3x3), ReLU \\ \cline{2-4}
                          & MaxPool 4   & 512 & MaxPool2d (2x2, stride=2) \\ \hline
Bottleneck               & ConvBlock 5 & 1024 & Conv2d (3x3), ReLU, Conv2d (3x3), ReLU \\ \hline
\multirow{4}{*}{Decoder} & UpConv 4    & 512 & ConvTranspose2d (2x2, stride=2) \\ \cline{2-4}
                          & ConvBlock 6 & 512 & Conv2d (3x3), ReLU, Conv2d (3x3), ReLU \\ \cline{2-4}
                          & UpConv 3    & 256 & ConvTranspose2d (2x2, stride=2) \\ \cline{2-4}
                          & ConvBlock 7 & 256 & Conv2d (3x3), ReLU, Conv2d (3x3), ReLU \\ \cline{2-4}
                          & UpConv 2    & 128 & ConvTranspose2d (2x2, stride=2) \\ \cline{2-4}
                          & ConvBlock 8 & 128 & Conv2d (3x3), ReLU, Conv2d (3x3), ReLU \\ \cline{2-4}
                          & UpConv 1    & 64 & ConvTranspose2d (2x2, stride=2) \\ \cline{2-4}
                          & ConvBlock 9 & 64 & Conv2d (3x3), ReLU, Conv2d (3x3), ReLU \\ \hline
Output Layer             & OutSeg      & 1 & Conv2d (1x1), Sigmoid \\ \hline
\end{tabular}%
}
\end{table}

The encoder is composed of four convolutional blocks, each containing two convolutional layers followed by ReLU activation functions. The number of feature maps doubles after each block, starting from 64 in the first block and increasing to 512 in the fourth block. After each convolutional block, a max-pooling operation with a kernel size of 2 and a stride of 2 is applied to downsample the feature maps, reducing their spatial dimensions while retaining important features. This progressive downsampling allows the encoder to extract hierarchical features from the input image.

The bottleneck layer serves as the bridge between the encoder and the decoder. It processes the feature maps at the lowest resolution using a convolutional block with 1024 feature maps. This layer captures the most abstract and high-level features of the input image, which are then passed to the decoder for reconstruction.

The decoder consists of four up-convolutional blocks, each of which uses a transposed convolutional layer (also known as a deconvolutional layer) to upsample the feature maps. The upsampled feature maps are concatenated with the corresponding feature maps from the encoder via skip connections, allowing the decoder to incorporate both high-level and fine-grained details. Each concatenation is followed by a convolutional block to refine the feature maps, gradually reducing their number while increasing their spatial dimensions. This symmetric structure ensures that the decoder can accurately reconstruct the spatial details of the input image.

The segmentation output is generated by a \(1 \times 1\) convolutional layer that reduces the number of feature maps to 1, followed by a sigmoid activation function. This output represents a probability map, where each pixel value indicates the likelihood of belonging to the tumor region. The U-Net architecture is designed to effectively capture both local and global features in medical images, making it well-suited for tumor segmentation tasks.

\subsection{Loss Function and Training}

The U-Net model is trained specifically for the task of image segmentation using a combined loss function that integrates Binary Cross-Entropy (BCE) Loss and Dice Loss. The training process is managed by a Trainer class, which orchestrates the training and validation loops, computes performance metrics, and saves the best-performing model. The key components of the training framework are described below.

\subsubsection{Loss Function}
The loss function used for training is a combination of Binary Cross-Entropy (BCE) Loss and Dice Loss, denoted as \(\mathcal{L}_{\text{total}}\). This combined loss function is defined as:

\[
\mathcal{L}_{\text{total}} = \mathcal{L}_{\text{BCE}} + \mathcal{L}_{\text{Dice}},
\]

where \(\mathcal{L}_{\text{BCE}}\) is the Binary Cross-Entropy Loss and \(\mathcal{L}_{\text{Dice}}\) is the Dice Loss. The Binary Cross-Entropy Loss is computed as:

\[
\mathcal{L}_{\text{BCE}} = -\frac{1}{N} \sum_{i=1}^{N} \left[ y_i \log(\hat{y}_i) + (1 - y_i) \log(1 - \hat{y}_i) \right],
\]

where \(y_i\) is the ground truth pixel value, \(\hat{y}_i\) is the predicted probability, and \(N\) is the total number of pixels. The Dice Loss, which measures the overlap between the predicted mask and the ground truth mask, is defined as:

\[
\mathcal{L}_{\text{Dice}} = 1 - \frac{2 \cdot |\text{Predicted} \cap \text{Target}| + \text{Smooth}}{|\text{Predicted}| + |\text{Target}| + \text{Smooth}},
\]

where \(\text{Smooth}\) is a small constant (typically \(1 \times 10^{-5}\)) added to avoid division by zero. The combined loss function ensures that the model is optimized for both pixel-wise accuracy and region-based overlap.

\subsubsection{Dice Coefficient}
To evaluate the segmentation performance, the Dice coefficient is employed as a metric. The Dice coefficient quantifies the overlap between the predicted binary mask and the ground truth mask. It is defined as:

\[
\text{Dice} = \frac{2 \cdot |\text{Predicted} \cap \text{Target}| + \text{Smooth}}{|\text{Predicted}| + |\text{Target}| + \text{Smooth}}.
\]

A higher Dice coefficient indicates better segmentation performance.

\subsubsection{Training and Validation}
The training process consists of two main loops: the training loop and the validation loop. During the training loop, the model is optimized using the combined loss function (\(\mathcal{L}_{\text{total}}\)). The validation loop evaluates the model's performance on a separate validation dataset, computing the Dice coefficient as the primary metric. Key metrics, including training loss, validation loss, and Dice coefficient, are logged and tracked throughout the training process.

\section{QuantU-Net}

QuantU-Net is an optimized version of the original U-Net architecture, designed to reduce computational complexity and memory usage by quantizing weights and activations to lower bitwidths. This quantization is achieved using the Brevitas library, which enables efficient deployment of deep learning models on resource-constrained devices. In this section, we describe the key differences between the QuantU-Net and the original U-Net.

QuantU-Net replaces standard layers in the original U-Net with their quantized Brevitas counterparts. Brevitas layers are designed to provide fine-grained control over quantization, allowing users to specify various parameters that determine how weights and activations are quantized. These parameters are critical for balancing model efficiency and accuracy. The main parameter that influences the quantization process is the bitwidth parameter. Lower bitwidths reduce memory usage and computational cost but may impact model accuracy. Higher bitwidths preserve accuracy but increase resource requirements.

In this paper, we define an optimal model as a QuantU-Net model where the average bitwidth is minimized, resulting in minimal resource usage while maintaining a negligible drop in accuracy (e.g., less than 2\%) compared to its floating-point counterpart. Given that our U-Net model consists of 23 convolutional layers and the bitwidth for each layer can vary between 2 and 8 bits, an exhaustive search for the optimal configuration would require evaluating \(6^{23}\) possible combinations, which is computationally infeasible. Instead, we opt to make the bitwidth itself a trainable parameter. This, combined with the loss function shown below, allows QuantU-Net to find the optimal model.


\subsection{QuantU-Net Loss Function}
The QuantU-Net extends the original U-Net implementation by introducing a bitwidth regularization loss to optimize the model for efficient inference on hardware with limited precision.
The QuantU-Net model employs a composite loss function that combines segmentation loss, Dice loss, and bitwidth regularization loss to optimize both model performance and quantization efficiency. The segmentation loss and the Dice loss are computed using the same methods as seen in the floating-point original. The QuantU-Net loss differentiates itself from its original by introducing a bitwidth loss function, which returns the average bitwidth over the entire model. When this function is used within the loss function, it penalizes higher bitwidths to encourage the use of lower precision weights and activations, thereby reducing model size and computational complexity. The total loss is formulated as:

\[
\mathcal{L}_{\text{total}} = \mathcal{L}_{\text{BCE}} +  \mathcal{L}_{\text{Dice}} + \lambda \cdot \mathcal{L}_{\text{Bitwidth}},
\]

where:

\begin{itemize}
    \item \(\mathcal{L}_{\text{BCE}}\) is the Binary Cross-Entropy (BCE) loss, computed as:
    \[
    \mathcal{L}_{\text{BCE}} = -\frac{1}{N} \sum_{i=1}^{N} \left[ y_i \log(\hat{y}_i) + (1 - y_i) \log(1 - \hat{y}_i) \right],
    \]
    where \(y_i\) is the ground truth pixel value, \(\hat{y}_i\) is the predicted probability, and \(N\) is the total number of pixels.
    
    \item \(\mathcal{L}_{\text{Dice}}\) is the Dice loss, which measures the overlap between the predicted mask and the ground truth mask:
    \[
    \mathcal{L}_{\text{Dice}} = 1 - \frac{2 \cdot |\text{Predicted} \cap \text{Target}| + \text{Smooth}}{|\text{Predicted}| + |\text{Target}| + \text{Smooth}},
    \]
    where \(\text{Smooth}\) is a small constant (typically \(1 \times 10^{-5}\)) added to avoid division by zero.
    
    \item \(\mathcal{L}_{\text{bitwidth}}\) is the bitwidth regularization loss, computed as:
    \[
    \mathcal{L}_{\text{Bitwidth}} = \text{Bitwidth},
    \]
    where \(\text{Bitwidth}\) is the average bitwidth of the model's quantized weights. 
    
    \item \(\lambda\) is a scaling factor set to 0.25 to balance the trade-off between segmentation accuracy and quantization efficiency.
\end{itemize}

During training, the model is optimized using gradient descent to minimize the total loss, ensuring that it achieves high segmentation accuracy while maintaining low bitwidth precision for efficient deployment on resource-constrained devices. The bitwidth regularization loss is particularly crucial for achieving a balance between model performance and hardware efficiency. As found in earlier works, initializing the model with a low bitwidth and having the loss function pull up the bitwidth if necessary results in a QuantU-Net model suitable for real-world wearable medical imaging applications.

\section{Experimental Set-up}

The experiments for this study were conducted in a Google Colab environment, which provides access to a Tesla T4 GPU for accelerated computation. The environment is equipped with a Tesla T4 GPU featuring 16 GB of GPU memory, enabling efficient training and inference for deep learning models. The CPU consists of 2 Intel(R) Xeon(R) processors running at 2.20 GHz, each with 1 core and 2 threads, providing a total of 2 logical processors. The system has approximately 12.68 GB of available RAM, with 7.63 GB free at the time of execution, and does not utilize swap memory as the swap size is set to 0 kB.

\section{Results}

The QuantU-Net model was trained over 40 epochs, and the results are summarized in Table~\ref{tab:quantnet_results}. The floating-point U-Net model achieved an accuracy of 96.14\%, serving as a strong baseline for comparison. QuantU-Net, with its quantized weights and activations, achieved a validation Dice coefficient of 0.4947 by the 10th epoch, demonstrating competitive segmentation performance despite the reduced precision. The average bitwidth across the model stabilized at approximately 4.24 bits, resulting in a model size that is $\sim$8 times smaller than the floating-point U-Net model.

\begin{table}[ht]
\centering
\caption{QuantU-Net Training Results}
\label{tab:quantnet_results}
\resizebox{\textwidth}{!}{%
\begin{tabular}{|c|c|c|c|c|c|}
\hline
\textbf{Epoch} & \textbf{Train Loss} & \textbf{Val Loss} & \textbf{Val Dice} & \textbf{Val Accuracy} & \textbf{Avg Bitwidth} \\ \hline
1  & 1.713  & 2.0265 & 0.1551 & 0.2845 & 4.0709 \\ \hline
10 & 0.9712 & 0.9727 & 0.4947 & 0.9421 & 4.2401 \\ \hline
20 & 1.1004 & 1.0587 & 0.2674 & 0.9349 & 4.1725 \\ \hline
30 & 0.9285 & 0.9493 & 0.4242 & 0.9417 & 4.5443 \\ \hline
40 & 0.9234 & 0.9601 & 0.3833 & 0.9399 & 4.5443 \\ \hline
\end{tabular}%
}
\end{table}

The validation accuracy of QuantU-Net remained consistently high, peaking at 94.25\% by the 39th epoch, which is only 1.89\% lower than the floating-point U-Net model's accuracy of 96.14\%. This minimal drop in accuracy is a remarkable achievement, considering the significant reduction in model size and computational complexity. The average bitwidth over the model increased slightly from 4.0709 bits in the first epoch to 4.5443 bits by the 30th epoch, reflecting the model's adaptation to maintain accuracy while optimizing for lower precision.

\begin{figure}[ht]
\centering
\includegraphics[width=0.8\textwidth]{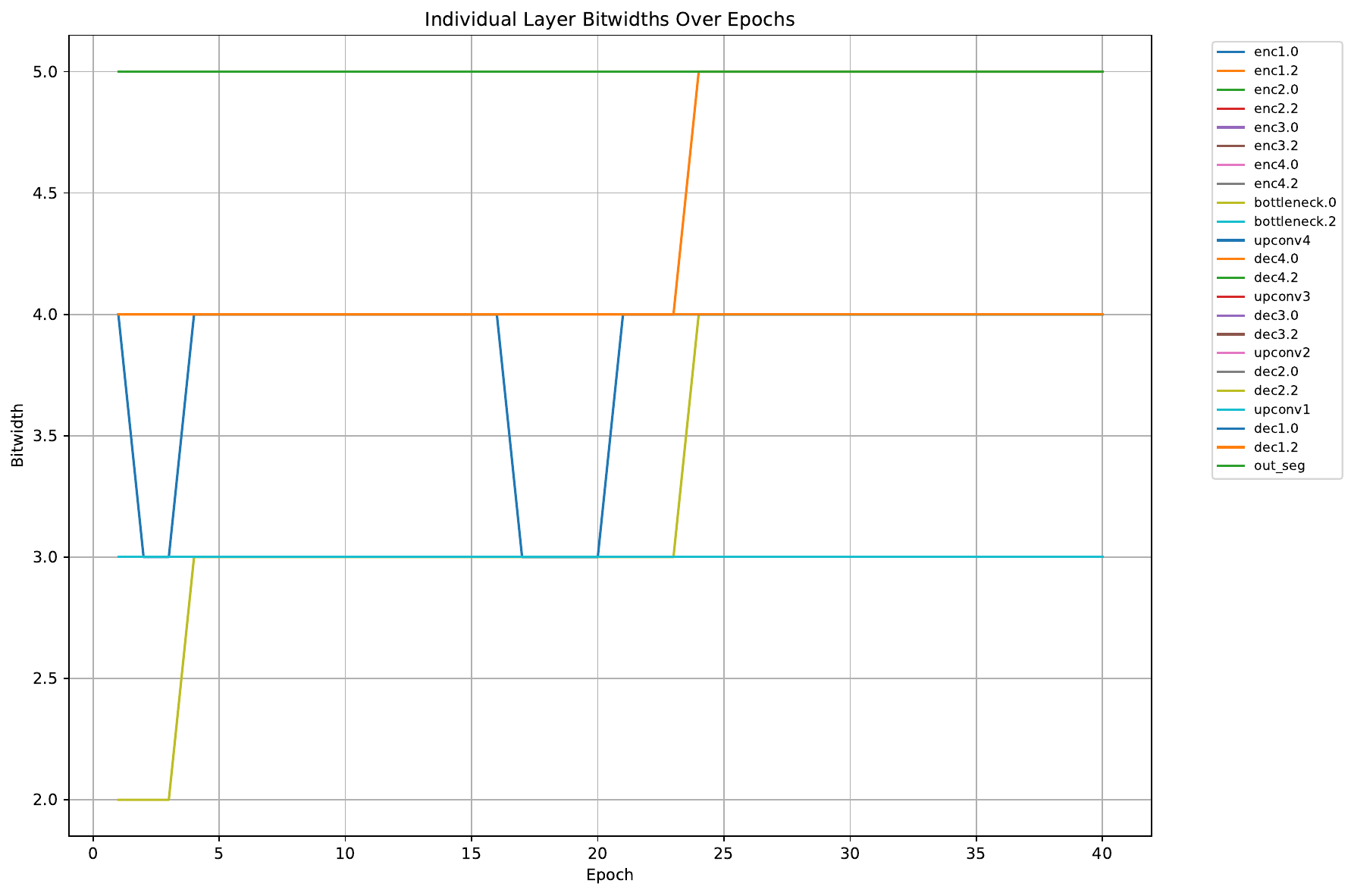}
\caption{Individual layer bitwidths over epochs. The plot shows how the bitwidths of different layers in the QuantU-Net model evolve during training. Layers such as the bottleneck and upconvolutional layers show dynamic adjustments to maintain accuracy while minimizing bitwidth.}
\label{fig:layer_bitwidths}
\end{figure}

The bitwidth distribution across the model layers remained stable, with most layers quantized to 4 or 5 bits. As shown in Figure~\ref{fig:layer_bitwidths}, the bitwidths of individual layers evolved dynamically during training. Layers such as the bottleneck and upconvolutional layers adjusted their bitwidths to maintain accuracy while minimizing precision. The bottleneck layer, which is critical for feature extraction, was quantized to 3 bits in the early epochs and increased to 4 bits in later epochs, indicating the model's prioritization of accuracy in critical layers.

\begin{figure}[ht]
\centering
\includegraphics[width=0.8\textwidth]{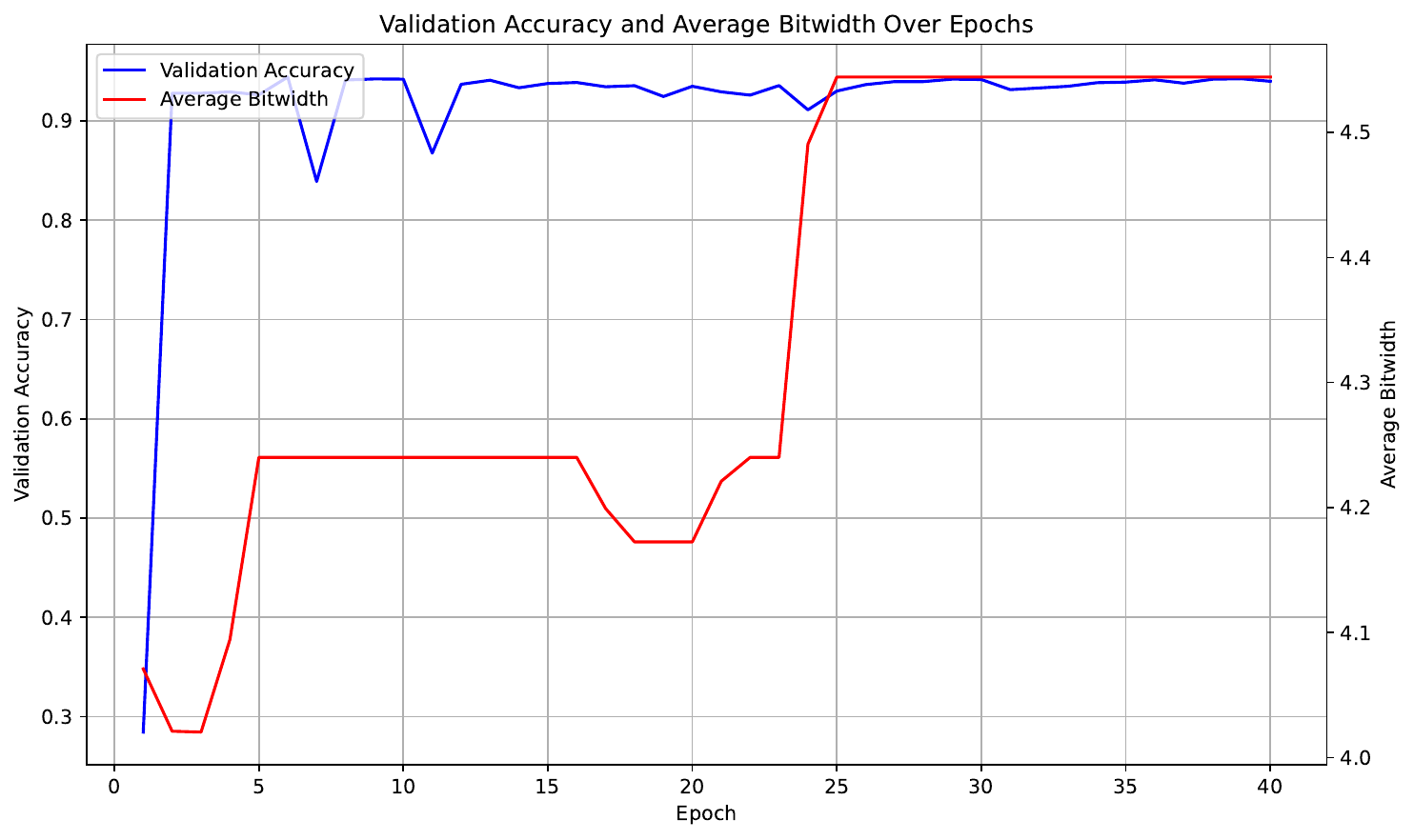}
\caption{Accuracy and average bitwidth over epochs. The left y-axis shows the validation accuracy, while the right y-axis shows the average bitwidth across the model. The plot demonstrates the trade-off between accuracy and bitwidth reduction, with accuracy remaining stable as the average bitwidth converges to around 4.24 bits.}
\label{fig:accuracy_bitwidth}
\end{figure}

Figure~\ref{fig:accuracy_bitwidth} illustrates the trade-off between validation accuracy and average bitwidth over epochs. The left y-axis shows the validation accuracy, while the right y-axis shows the average bitwidth across the model. The plot demonstrates that accuracy remained stable as the average bitwidth converged to around 4.24 bits, highlighting the effectiveness of the quantization approach.

\begin{figure}[ht]
\centering
\includegraphics[width=0.8\textwidth]{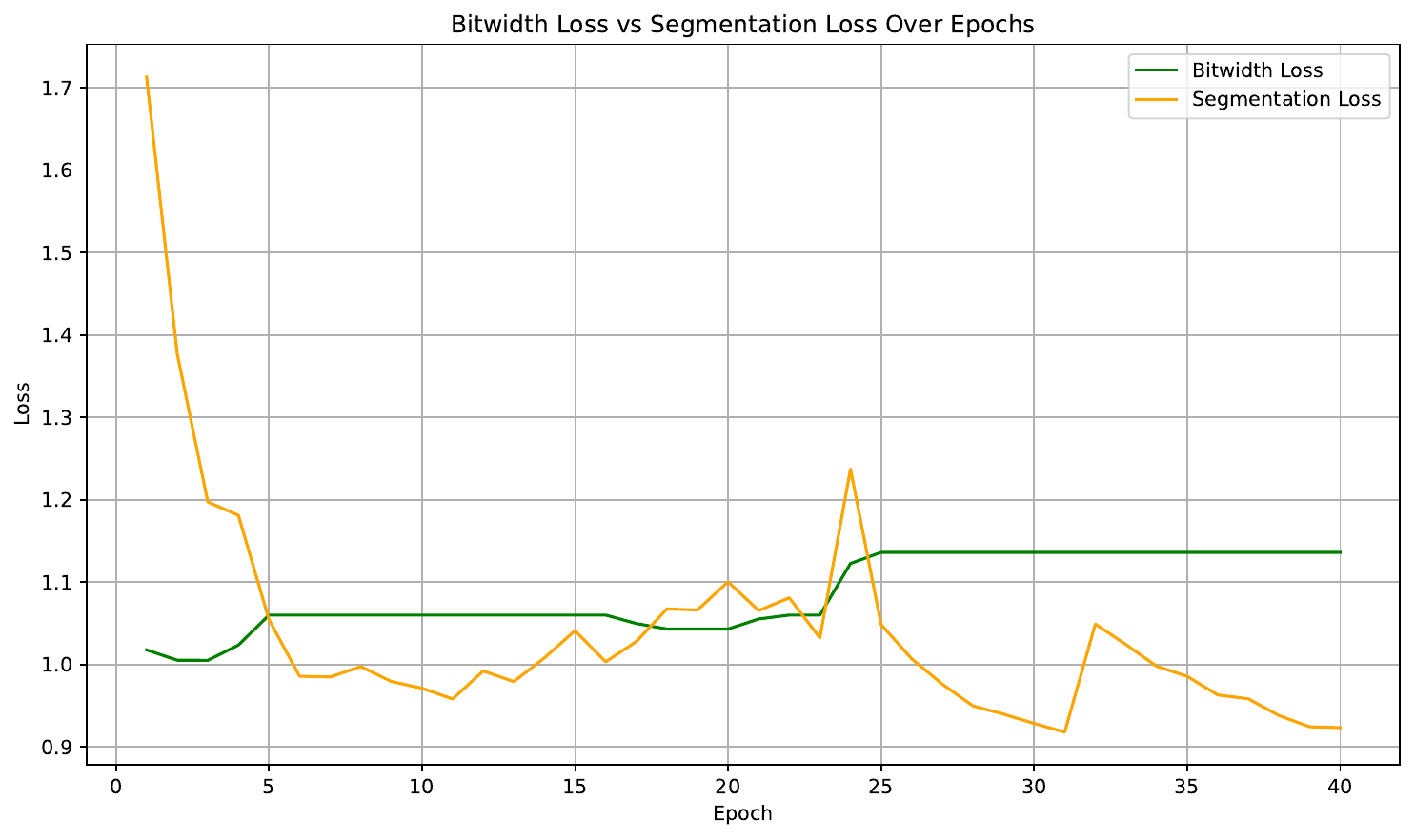}
\caption{Bitwidth loss vs segmentation loss. The plot illustrates the relationship between the bitwidth regularization loss and the segmentation loss over training epochs. The bitwidth loss decreases as the model optimizes for lower precision, while the segmentation loss remains stable, indicating that the model maintains accuracy despite the reduction in bitwidth.}
\label{fig:bitwidth_vs_segmentation}
\end{figure}

Additionally, Figure~\ref{fig:bitwidth_vs_segmentation} shows the relationship between the bitwidth regularization loss and the segmentation loss over training epochs. The bitwidth loss decreased as the model optimized for lower precision, while the segmentation loss remained stable, indicating that the model maintained accuracy despite the reduction in bitwidth.

All of these results were achieved within a single training session, in contrast to the hypothetical \(6^{23}\) training sessions that would be required to find the optimal bitwidth through exhaustive search. This efficiency underscores the effectiveness of the proposed bitwidth optimization approach, which leverages gradient descent and a custom loss function to dynamically determine the smallest possible bitwidth without sacrificing accuracy.

\section{Discussion}

The results demonstrate that QuantU-Net successfully balances the trade-off between model efficiency and segmentation accuracy. While the floating-point U-Net model achieved an accuracy of 96.14\%, QuantU-Net achieved a validation accuracy of 94.25\%, with a model size that is $\sim$8 times smaller. This reduction in model size is a direct result of quantizing the weights and activations to an average of 4.24 bits, significantly reducing memory usage and computational cost.

The use of integer operations in QuantU-Net, as opposed to floating-point operations in the baseline model, opens up significant opportunities for deployment on FPGAs and other AI accelerators. FPGAs are particularly well-suited for running quantized models due to their ability to perform fixed-point arithmetic more efficiently than floating-point operations. 

The minimal drop in accuracy (from 96.14\% to 94.25\%) is a testament to the effectiveness of Quantization Aware Training (QAT) in maintaining model performance while reducing precision. The bitwidth regularization loss played a crucial role in guiding the model to use lower bitwidths where possible, without sacrificing accuracy. This is particularly evident in the bottleneck layer, where the bitwidth increased slightly in later epochs to maintain feature extraction quality.

One notable observation is the fluctuation in the validation Dice coefficient, which peaked at 0.4947 in the 10th epoch but varied in subsequent epochs. This variability suggests that while the model achieved high accuracy, there is still room for improvement in terms of stability. Future work could explore adaptive quantization techniques that dynamically adjust bitwidths based on the complexity of the input data, potentially improving both accuracy and stability.

The reduction in model size and the use of integer operations make QuantU-Net highly suitable for deployment on resource-constrained devices, such as wearable medical devices and embedded systems. The potential efficiency gains on FPGAs and other AI accelerators further enhance the model's applicability in real-world scenarios, where low power consumption and real-time performance are critical.

\section{Conclusion}

This study presented a comprehensive analysis of the U-Net architecture for tumor segmentation, addressing the computational challenges associated with deploying deep learning models on resource-constrained devices. By leveraging Quantization Aware Training (QAT) and utilizing Brevitas for bitwidth optimization, the QuantU-Net model achieved a significant reduction in computational complexity and memory usage while maintaining competitive segmentation accuracy.

The results demonstrate that QuantU-Net effectively balances the trade-off between model efficiency and accuracy, making it suitable for deployment in wearable medical devices and embedded systems. The bitwidth optimization approach not only reduced model size but also opens the way to enhance energy efficiency which is crucial for continuous health monitoring applications.

Despite these advancements, challenges remain, particularly in balancing quantization efficiency with minimal accuracy degradation. Future work could explore adaptive quantization techniques and hybrid models to further optimize performance. Additionally, validating the QuantU-Net model on larger, more diverse datasets and deploying it on various hardware platforms—such as ASICs and other neural network accelerators—to assess latency, power consumption, and memory efficiency would enhance its applicability in real-world clinical scenarios.

This study contributes to the field of medical image segmentation by demonstrating the potential of quantization-aware U-Net models for efficient and accurate tumor segmentation, paving the way for real-time, low-power, and wearable diagnostic solutions.

\section{Acknowledgements}

We thank Giuseppe Franco for his assistance with bug fixing the Brevitas code, it has greatly helped with this work.

\bibliographystyle{plain} 
\bibliography{refs} 

\end{document}